\documentstyle[12pt,aps,prb,floats,eqsecnum,psfig]{revtex}
\def\lapprox{\,\raise0.4ex\hbox{$<$}\kern-0.8em\lower0.7ex\hbox{$\sim$}\,}
\def\gapprox{\,\raise0.4ex\hbox{$>$}\kern-0.8em\lower0.7ex\hbox{$\sim$}\,}

\begin{document}
\bibliographystyle{prsty}
\centerline{ \bf SOUND AND HEAT ABSORPTION BY A 2D ELECTRON GAS }
\centerline{\bf IN AN ODD-INTEGER QUANTIZED-HALL REGIME }

$\quad $

\centerline{S. Dickmann\footnote{e-mail: dickmann@issp.ac.ru}}

\centerline{\it Max Planck Institute for Physics of Complex Systems,}

\centerline{\it N\"othnitzer Str. 38, D-01187 Dresden;}

\centerline{\it Institute for Solid State Physics of Russian Academy of
Sciences,}

\centerline{\it 142432 Chernogolovka, Moscow District, Russia}

\vskip 8mm

\begin{abstract}
The absorption of bulk acoustic phonons in a two-dimensional (2D)
GaAs/AlGaAs heterostructure is studied
(in the clean limit) where the 2D electron-gas (2DEG), being
in an odd-integer quantum-Hall state, is in fact a spin dielectric. Of the
two channels of phonon absorption associated with excitation of spin waves,
one, which is due to the spin-orbit (SO) coupling of electrons,
involves
a change of the spin state of the system and the
other does not. We show
that the phonon-absorption rate corresponding to the former channel
(in the paper designated as the second absorption channel)
is finite at zero temperature ($T$), whereas that corresponding to
the latter (designated as the first channel) vanishes for $T\to 0$.
The long-wavelength limit, being the special case of the first
absorption channel, corresponds to sound (bulk and surface)
attenuation by the 2DEG. At the same time, the ballistic phonon
propagation and heat absorption are determined by both
channels. The 2DEG overheat and the attendant spin-state change are
found under the conditions of permanent nonequilibrium phonon
pumping.

\end{abstract}

\begin{flushleft}
${}\qquad{}\qquad{}$ PACS: 73.40.Hm, 63.22.+m, 71.70.Ej, 75.30.Fv
\end{flushleft}

\section{Introduction}
In recent years considerable amount of interest has been focused on the 
problem of
the acoustic wave and heat absorption by a 2DEG in
GaAs/AlGaAs-heterostructures in the quantum Hall regime (QHR)
\cite{ei87,ch90,ke92,ra92,wi86,wi89,es94,wi90,wi94,io89,io96,kn96}. This is
connected with the search for a new way to study the fundamental properties
of a 2DEG in a strong magnetic field (which is considered to be
perpendicular to the layer, i.e., $\mbox{\boldmath $B$}\parallel{\hat z}$)
employing nonequilibrium phonons \cite{ch90,ke92} or surface acoustic waves
\cite{ra92,wi86,wi89,es94,wi90,wi94} as an experimental tool. The basic idea
is associated with the fact that the energies of phonons generated by
heated metal films \cite{ch90,ke92,we86} or the energy of coherent phonons
in semiconductor superlattices \cite{ba99} may be of the order of
the characteristic gaps in the 2DEG spectrum. At the same time, it is
well-known that in the QHR a change of the Landau level (LL) filling factor 
$\nu$ may drastically renormalize the 2DEG
excitation spectrum. Therefore parameters such as the phonon life-time
(PLT) $\tau_{\mbox{\scriptsize ph}}$, the
attenuation, and the velocity of sound waves exhibit strong oscillations as
functions of the applied magnetic field
\cite{ei87,ra92,wi86,wi89,es94,wi90,wi94}. These spectrum
alterations prevent development of an universal approach to the theoretical
problem of sound and heat absorption by 2DEG.

Most of the relevant treatments\cite{io89,io96,kn96,al94} use
the one-particle approximation, i.e. the Coulomb electron--electron
($e$-$e$) interaction is
neglected or considered as a secondary phenomenon renormalizing the phonon
displacement field (screening in Ref. \onlinecite{kn96}) or one-electron
state density (Coulomb gap in Ref. \onlinecite{al94}). In these studies the
LL width is determined by the amplitude $\Delta$ of the smooth random
potential (SRP), and the phonon absorption occurs through the transition
of an
electron from one semiclassical trajectory to another in the SRP field
near the percolation threshold\cite{io96,kn96} (when $\nu$ is close to
a half-integer) or by the electron variable-range hopping for strongly
localized electrons\cite{al94} ($\nu$ is close to an integer). The
one-particle
approximation is evidently justified in a qualitative sense as long as
the number
of charged quasiparticles is rather large, which does not hold in the
integer (or ``almost" integer) QHR. In this latter
case one should take into account a strong $e$-$e$ interaction with the
typical energy equal to the Coulomb energy $E_C\sim
e^2/\varepsilon\l_B$
($\varepsilon$ is the dielectric constant, $l_B=(\hbar c/eB)^{1/2}$ is the
magnetic
length), which exceeds {100\,}K for $B > 8\,$T. Since we have $E_C/\Delta
\gapprox 10$ and because of the absence of charged quasiparticles in the
ground state,
it is exactly $E_C$ that determines the real LL width in the integer QHR.
Charge excitations are separated from the ground state, not only by the
gap, which is equal to the Zeeman energy for odd $\nu$ or the
cyclotron energy $\hbar\omega_c$ for even $\nu$, but also by
an additional energy of order $E_C$. This applies also
to the so-called skyrmion charge
excitations\cite{so93,fe94,br95}, which seem to have been
observed\cite{ba95,ku97}
at $\nu=1$. However, if $\nu$ deviates from unity then even the ground
state has to be realized as a complex spin and charge texture in the form of a
Skyrme crystal with a characteristic period proportional to ${|\nu
-1|}^{-1/2}$ (see Refs. \onlinecite{so93,by96,io97}), so that ignoring the
skyrmions is only permissible for $\nu$ close enough to unity. 
Experiments\cite{ba95,ku97}
indicate that this can indeed be done if $|\nu -1|\lapprox 0.01$.

Therefore in integer QHR the phonon absorption by chargeless excitations has
to be more efficient. The spectrum calculation for low energy excitations
from the filled LL is fortunately an exactly solvable problem to first
order in $E_C/\hbar\omega_c$, which is considered to be small
(see Refs. \onlinecite{by81,ka84}). We will study the odd $\nu$ only, since
at even $\nu$
the cyclotron gap for excitations substantially exceeds the possible
acoustic phonon energy. For $\nu=2n+1$ when the $n$-th LL
in the ground state has a fully occupied lower spin sublevel and an
empty upper one the lowest states in the spectrum are spin
excitons, which are in fact spin wave excitations.
We will call these simply spin waves (SWs). For them the gap
is $|g\mu_bB|\approx 0.3B\,$K/T,
because $g=-0.44$ for GaAs. At temperatures $T\gapprox 0.1\,$ an
appreciable amount of ``thermal" SWs forms
a rarefied 2D Bose gas, since SW density is still far less than the
density of electrons in the $n$-th LL (i.e., than $1/2\pi l_B^2$).
As a result, the electron--phonon ($e$-$ph$) interaction can be
represented as the SW--phonon
interaction. Such a representation for the $e$-$ph$ interaction
Hamiltonian has already been found earlier\cite{di96} (see also Sec. 2
of the present
paper), and as discussed it includes, in addition to spin-independent
terms, the small terms
arising from the electron SO coupling. Precisely the
latter ones determine the phonon absorption at $T=0$.

We consider two channels of inelastic phonon scattering. The first one
(Sec. 3 of the present paper) does not change the 2DEG spin state. The rate
of phonon absorption is proportional
to the conserved number of equilibrium SWs and vanishes when $T$ goes
to zero. Therefore, as the 2DEG temperature due to phonon absorption
increases the SWs chemical potential has to decrease in this case.
After averaging with a certain phonon distribution
over the momenta we find the
mean effective inverse PLT $1/\tau_{\mbox{\scriptsize eff}}$, and hence the
rate of the 2DEG heating as well as the corresponding contribution to the
inverse thermal conductivity. The transition to the limit $kl_B \ll 1$
($\mbox{\boldmath $k$}=(\mbox{\boldmath $q$},k_z)$ is the phonon wave
vector) allows one to find the ``2D" contribution to the bulk and surface
acoustic wave attenuation. In this last case the piezoelectric
electron--lattice interaction plays the main role.

The second phonon absorption channel (Sec. 4) arises from the SO
coupling. It describes the SW generation from the 2DEG ground
state. Absorption of a single phonon reduces by 1 both the spin
component $S_z$
and the total electron-spin number $S$. As a consequence, the
absorption rate is proportional to the rate of
spin momentum decrease. This channel of scattering which is pinched off
for energies less than the Zeeman energy gap, is only accessible for
a selected group of ``resonant" phonons with a certain kinematic
relationship between
$q$ and $k_z$ wave vector components. Of course, the long-range wave
limit is meaningless in this case. While the phonons
of the resonant group occupy a comparatively small phase volume in
$k$-space, their contribution to the effective inverse PLT is rather
significant and, being independent of temperature, can exceed  the corresponding contribution from the first
absorption channel even at $T\lapprox 1\,$K.

In Section 5 we consider both absorption channels in the
problem of dynamic quasi-equilibrium in a 2DEG under the condition of
ballistic phonon pumping. We find the dependence of the final temperature 
as well as the spin momentum of the 2DEG on the initial temperature
of the 2DEG and the density of the nonequilibrium phonons.
We note in passing that the 2DEG adds a small correction to the bulk phonon 
scattering, connected mainly with the lattice defects and sample boundaries.
One can also get only a small (although peculiar) 2DEG correction to
the thermal conductance\cite{ei87,io89}. The 2DEG contribution to the
phonon absorption always contains the factor $1/L_z$ ($L_z$ is the sample
thickness in the ${\hat z}$ direction) in the expression
for the value of $1/\tau_{\mbox{\scriptsize ph}}$. We assume that the 
distribution of the nonequilibrium phonons in $k$-space and the amplitude
of the sound wave in the 2D channel, 
which are determined by the acoustic-phonon scattering, to be known. 

In closing this Section we should like to mention the SRP role
specifically in the case of the studied problem.
The ground state in the clean limit with strictly odd $\nu$, being built of
one-electron states of the fully occupied lower spin sublevel
(see Refs. \onlinecite{by81,ka84,di96,by83}), is actually the
same as for the 2DEG without Coulomb interaction. Accordingly, the ground 
state in the presence of SRP is of the same nature. 
Note that the SRP would give rise to free charged 
quasiparticles, were it not that the loss in energy due to interaction 
(which is on the order of $\vert g \mu_b B\vert + E_C$), was vastly
in excess of the gain $\Delta$ in energy due to fluctuations in the random 
potential\cite{foot1}. 
We have no free quasiparticles at temperatures $T
\ll E_C$, and such a 2DEG is a spin dielectric.
Furthermore the neutral spin exciton with 2D
momentum $\hbar{\mbox{\boldmath $q$}}_{\mbox{\scriptsize ex}}$ has a dipole 
momentum
$el_B^2\mbox{\boldmath $q$}_{\mbox{\scriptsize ex}}\times {\hat z}$ (see Refs.
\onlinecite{by81,ka84,le80}) and in
SRP the exciton may gain an energy on order
$l_B^2q_{\mbox{\scriptsize ex}}\Delta/\Lambda$ in the dipole approximation, 
where $\Lambda$ is SRP correlation length. The
associated loss is the ``kinetic" energy $(q_{\mbox{\scriptsize ex}}l_B)^2/2M$
($M\sim E_C^{-1}$ is the excitonic mass\cite{by81,ka84,le80}). Therefore for
$q_{\mbox{\scriptsize ex}}\lapprox
q_0=\Lambda^{-1}\Delta M\sim \Delta/E_C\Lambda$ one has to
take into account the SRP effect on the SW energy. Our
approach to the phonon absorption by the equilibrium SWs ignores the
presence of SRP in the system, so it is only correct for temperatures
$T> (q_0l_B)^2/2M \sim 10\,$mK (this estimate has been done for $B=10\,$T,
$\Delta=1\,$meV, and $\Lambda=50\,$nm).

\section{One-exciton states and the electron--phonon
Interaction Hamiltonian in the Excitonic Representation}
In this Section we introduce the so-called `excitonic representation' of the
Hamiltonian, and its eigenstates, describing the 2DEG
under consideration. Let $a_p$ and
$b_p$ be annihilation
operators for an electron in the Landau gauge state $\Psi_p(x,y)=
L^{-1/2}e^{ipy}\psi_n(pl_B^2+x)$
at the lower and upper spin sublevels, respectively. Here $L\times L$ is the
size of the 2D system, and $\psi_n$ is the $n$-th harmonic oscillator
function. In the ``one-exciton''approximation, the absorption of one phonon 
which is not accompanied by a change in the spin state of the 2DEG
amounts to a transition between the one-spin-wave states of the 2DEG.
The one-spin-wave state with 2D
momentum $\hbar\mbox{\boldmath $q$}$ is defined as
$$
  |\mbox{\boldmath $q$}\rangle=Q^+_{\bf q}|0\rangle\,.  \eqno(2.1)
$$
Here the creation operator for SW,
$$
  Q_{\bf q}^{+}={\cal N}^{-1/2}\sum_{p}\,
  e^{-iq_x p}
  b_{p+\frac{q_y}{2}}^{+}\,a_{p-\frac{q_y}{2}}\,,    \eqno (2.2)
$$
operates on the 2DEG ground state $|0\rangle$, and
${\cal N}={L^2}/{2\pi l_B^2}$ is the number of magnetic flux quanta, or
equivalently, the number of electrons in the $n$-th LL. 
The equations
$$
  a^+_p|0\rangle=b_p|0\rangle\equiv 0\quad \mbox{for any $p$},
                                                            \eqno (2.3)
$$
may be regarded as the definition for $|0\rangle$. The main aspect of
the excitonic representation is that the state (2.1) is an eigenstate
of the
full electron Hamiltonian ${\cal H}$ involving the $e$-$e$ interaction:
${\cal H}|\mbox{\boldmath $q$}_{\mbox{\scriptsize ex}}\rangle=
[E_0+\epsilon(\mbox{\boldmath $q$}_{\mbox{\scriptsize ex}})]
|\mbox{\boldmath $q$}_{\mbox{\scriptsize ex}}\rangle$
($E_0$ is the 2DEG
ground state energy and
$\epsilon(\mbox{\boldmath $q$}_{\mbox{\scriptsize ex}})$ is the SW
energy). Of course, this
is valid to first order in $E_c/\hbar\omega_c$ and in the 2D
limit (we assume that $d<l_B<r_b$, where $r_b$ is the Bohr radius in the
material and $d$ is the effective 2DEG thickness). In the
limit $q_{\mbox{\scriptsize ex}}l_B \ll 1$ appropriate to our problem
one may use the quadratic approximation for the excitonic energy,
$$
  \epsilon(q_{\mbox{\scriptsize ex}})=
  \delta + \frac{(q_{\mbox{\scriptsize ex}}l_B)^2}{2M}\,,    \eqno (2.4)
$$
where
$$
  \delta =|g\mu_bB|\,,                      \eqno (2.5)
$$
and
$$
  \quad M^{-1}=
  l_B^2\int_{0}^\infty \frac{p^3dp}{4\pi}V(p)
  e^{-p^2l_B^2/2}[L_n(p^2l_B^2/2)]^2                    \eqno (2.6)
$$
($L_n$ is the $n$-th order Laguerre polynomial), which is defined in terms of the Fourier
components $V(q)$ of the Coulomb potential in the heterostructure averaged
over the 2DEG thickness\cite{di96,by83}. Note that $1/M\sim E_c$ holds,
besides  our results depend on the LL number $n$ only because the excitonic
mass (2.6) depends on $n$.

The excitonic representation for the interacting Hamiltonian involves the 
displacement operators:
$$
{A_{\bf q}^{+}\choose B_{\bf q}^{+}}=
{A_{-{\bf q}}\choose B_{-{\bf q}}}=
\frac{1}{\cal N}\sum_{p}
e^{-iq_{x}p}{a_{p+\frac{q_{y}}{2}}^{+}
a_{p-\frac{q_{y}}{2}}\choose
b_{p+\frac{q_{y}}{2}}^{+}b_{p-\frac{q_{y}}{2}}}\,.  \eqno (2.7)
$$
The identities in (2.3) can be rewritten in the excitonic representation
as
$$
  A^+_{\bf q}|0\rangle\equiv \delta_{0,{\bf q}}|0\rangle,\quad
  B^+_{\bf q}|0\rangle=Q_{\bf q}|0\rangle\equiv 0\,.
                                              \eqno (2.8)
$$

The operator in Eq. (2.2) (as well as its Hermitian conjugate) seems to have 
been first introduced in
Ref. \onlinecite{dz84}. Later some of its combinations together with the
operators (2.7)
were in fact used as the ``valley wave"\cite{ras86,by87} or
``iso-spin"\cite{by96} operators. In Ref. \onlinecite{di96} the operators
(2.2) and (2.7) have been considered exactly in the present form. In what 
follows, we shall take advantage of the commutation relations
$$
e^{-i\Theta_{12}}[A_{{\bf q}_1}^+,Q^+_{{\bf q}_2}]=
  -e^{i\Theta_{12}}[B_{{\bf q}_1}^+,Q^+_{{\bf q}_2}]=
  -{\cal N}^{-1}Q^+_{{\bf q}_1\!+\!{\bf q}_2}\,,    \eqno (2.9)
$$
and
$$
[Q_{{\bf q}_1},Q_{{\bf q}_2}^{+}]=
  e^{i\Theta_{12}}A_{{\bf q}_1\!-\!{\bf q}_2}-
  e^{-i\Theta_{12}}B_{{\bf q}_1\!-\!{\bf q}_2}\,,  \eqno (2.10)
$$
where $\Theta_{12}=
  l_B^2(\mbox{\boldmath $q$}_1\times \mbox{\boldmath $q$}_2)_z/2$.

The $e$-$ph$ interaction Hamiltonian is presented as
(see, e.g., Refs.\onlinecite{io89,di96}):
$$
  {\cal H}_{\mbox{\scriptsize e,ph}}=\frac{1}{L}\left(\frac{\hbar}{L_z}
  \right)^{1/2}\,\sum_{{\bf q},{k}_z,s}
  {U'}_{s}(\mbox{\boldmath $k$})\,
  P_{{\bf k},s}
  H_{\mbox{\scriptsize e,ph}}(\mbox{\boldmath $q$})\qquad+\qquad
  \mbox{H.~c.}\,,\eqno (2.11)
$$
where $P_{{\bf k},s}$ is the phonon annihilation operator (index $s$ denotes
the polarization state, with $s=l$ denoting the longitudinal and $s=t$ the
transversal polarization state). The $H_{\mbox{\scriptsize e,ph}}$ operates on 
the electron states, and ${U'}_{s}(\mbox{\boldmath $k$})$ is the renormalized
vertex which includes the fields of deformation (DA) and piezoelectric
(PA) interactions. The integration with respect to $z$ has been already
performed, and reduces to the renormalization
${ U'}_{s}(\mbox{\boldmath $k$})=
\gamma(k_z)U_{s}(\mbox{\boldmath $k$})$, where the formfactor
$$
  \gamma(k_z)=\int
  f^{*}(z)e^{ik_{z}z}f(z)\,dz                       \eqno (2.12)
$$
is determined by the wave function $f(z)$ of the corresponding
size-quantized level (which we have assumed to be identical for all ${\cal N}$
electrons). For the
three-dimensional (3D) vertex one needs only the expressions for the squares,
$$
  |U_s|^2=\pi\hbar\omega_{s,k}/p_0^3\tau_{A,s}(\mbox{\boldmath $k$})\,,
  \eqno (2.13)
$$
where $\omega_{s,k}=c_sk$ are the phonon frequencies, $p_0=2.52\cdot
10^6\,$cm${}^{-1}$ is the material parameter of GaAs (Ref.
\onlinecite{gale87}), and
$c_l$ and $c_t$ are the sound velocities.
The longitudinal $\tau_{A,l}(\mbox{\boldmath $k$})$ and transverse
$\tau_{A,t}(\mbox{\boldmath $k$})$ times are 3D acoustic
phonon life-times (see Appendix~I). These quantities
are expressed in terms of nominal times $\tau_D=0.8\,$ps and
$\tau_P=35\,$ps characterizing respectively DA and PA phonon
scattering in three-dimensional GaAs crystal.
(See Ref. \onlinecite{di96} and cf. Ref. \onlinecite{gale87}.)

Initially, of course, the originally spin-independent Hamiltonian of the
$e$-$ph$ interaction (2.11) is used. However it does not commute with the
SO coupling part of the electron Hamiltonian. Therefore the operator
$H_{\mbox{\scriptsize e,ph}}$
including the relevant off-diagonal corrections in the excitonic
representation has the following form\cite{di96} (we write it for
$ql_B\ll 1$):
$$
  H_{\mbox{\scriptsize e,ph}}(\mbox{\boldmath $q$})=
  \left\{ {\cal N}(A_{\bf q}^+ + B_{\bf q}^+)\right.\quad{}\qquad{}\qquad{}
$$
$$
  {}\qquad{}\qquad{}+\left.{\cal N}^{1/2}l_B\left[ (vq_-- i uq_{+})
  Q_{-{\bf q}} - (i uq_-+vq_{+})Q_{\bf q}^+\right] \right\}\,,
                                                      \eqno (2.14)
$$
where $q_{\pm}=\mp i 2^{-1/2}(q_x\pm iq_y)$.
Here $u$ and $v$ are dimensionless
parameters (just as in Ref. \onlinecite{di96}) characterizing the
SO coupling. The terms containing the coefficient $v$
originate from the absence of inversion
in the direction perpendicular to the 2D layer and hence
$v$ is proportional to the strength of the normal electric field in the
heterostructure\cite{by84}. The terms including $u$ are related to
the lifting of spin degeneracy for the $S$-band in A${}_3$B${}_5$
crystals\cite{d'86}. In deriving Eq. (2.14), we have assumed
that the normal ${\hat z}$ is parallel to the principal [001] axis of
the crystal. The final results depend only on the
combination $u^2+v^2$, which is of
order $10^{-5}$ for $B=10\,$T with $d=5\,$nm and is proportional
to $B^{-1}$ and also to $d^{-4}$ in the asymptotic limit $d\to 0$.

Further in our estimations we proceed from
the fields $B\simeq 5\div 20\,$T, and so $\delta$ (2.5) has the same
order of magnitude as $\hbar c_s/l_B$
(precisely $l_B\simeq 6\div 11\,$nm,${}\quad\delta \simeq 1.5\div 6\,\mbox{K},
\quad \hbar
c_ll_B^{-1} \simeq\ 3\div 7\,\mbox{K},\quad \hbar c_tl_B^{-1}\simeq 2\div
4\,\mbox{K}$). The magnitude of the exciton mass depends on the layer
thickness
according to Eq. (2.6) because $V(q)$ depends on the size-quantized
function $f(z)$; for real heterostructures $M^{-1}\simeq 40\div 80\,$K.

Note also that everywhere below, the specific magnetic-field dependence
of our results is given at constant $\nu$, i.e., the surface electron
density is always considered to be proportional to $B$.

\section{Phonon absorption without a change of the spin state (the first
absorption channel)}
Considering for the present the first absorption channel, we
find the PLT from the well known formula
$$
  \frac{1}{\tau_s(\mbox{\boldmath $k$})}=\sum_{i\ne f}
  W_{if}(\mbox{\boldmath
  $k$})[b_T(\epsilon_i)-b_T(\epsilon_f)],        \eqno (3.1)
$$
where the scattering probability
$$
  W_{if}(\mbox{\boldmath $k$})=\frac{2\pi}{\hbar}|{\cal
  M}_{if}^{s,{\bf k}}|^2\delta(\epsilon_i-
  \epsilon_f+\hbar\omega_{s,k})              \eqno (3.2)
$$
contains the matrix element of the Hamiltonian (2.11) (Ref.
\onlinecite{foot2}). The value of ${\cal M}_{if}^{s,{\bf k}}$ is
determined by the annihilation of one
phonon having momentum $\hbar\mbox{\boldmath $k$}$ and by the
transition of the 2DEG between the states
$|i\rangle=|\mbox{\boldmath $q$}_{\mbox{\scriptsize ex}}\rangle$ and
$\langle f|=\langle\mbox{\boldmath $q$}_{\mbox{\scriptsize ex}}
\!+\!\mbox{\boldmath $q$}|$,
where $\mbox{\boldmath $q$}$
is the component of $\mbox{\boldmath $k$}$ in the 2DEG plane
(for the analogous formulae for phonon absorption by free electrons,
see, e.g., Refs. \onlinecite{io89,kn96,gale87}). Here
$\epsilon_i=\epsilon(\mbox{\boldmath $q$}_{\mbox{\scriptsize ex}})$ and
$\epsilon_f=\epsilon
(\mbox{\boldmath $q$}_{\mbox{\scriptsize ex}}+\mbox{\boldmath $q$})$
and the function $b_T(\epsilon)$ in Eq. (3.1) corresponds to the Bose
distribution for SWs,
$$
  b_T(\epsilon)=\frac{1}{\exp{[(\epsilon-\mu)/T]}-1}\,,\qquad
  \mbox{where}\qquad \mu < \delta\,.  \eqno (3.3)
$$
According to Eqs. (2.11) and (2.14) the square of the modulus
is
$$
  |{\cal
  M}_{if}^{s,{\bf k}}|^2=\frac{\pi|\gamma(k_z)|^2\hbar^2\omega_{s,k}}
  {p_0^3L^2L_z\tau_{A,s}(\mbox{\boldmath $k$})} \left|\langle
  \mbox{\boldmath
  $q$}_{\mbox{\scriptsize ex}}+\mbox{\boldmath $q$}|
  H_{\mbox{\scriptsize e,ph}} (\mbox{\boldmath
  $q$})|\mbox{\boldmath $q$}_{\mbox{\scriptsize ex}}\rangle\right|^2 \,.
                                                         \eqno (3.4)
$$
Naturally, we suppose
that the internal equilibrium in the 2DEG among SWs is
established faster than the equilibrium between the phonons and the 2DEG
(see the comment in Appendix~II).

Equating the argument of the $\delta$-function in Eq. (3.2)
to zero,  $(q_{\mbox{\scriptsize ex}}l_B)2/2M+\hbar c_s k=
|\mbox{\boldmath $q$}_{\mbox{\scriptsize ex}}+
\mbox{\boldmath $q$}|^2l_B^2/2M$, one can find
the kinematic relationship for
$\mbox{\boldmath $q$}_{\mbox{\scriptsize ex}}$, which reduces to
$$
  q_{\mbox{\scriptsize ex}}\cos{\beta}=
  R_{s,{\bf k}}=\hbar c_s l^{-2}_BMk/q-q/2\,,
                                               \eqno (3.5)
$$
where $\beta$ is the angle between
$\mbox{\boldmath $q$}_{\mbox{\scriptsize ex}}$
and \mbox{\boldmath $q$}. We have used here the quadratic
approximation (2.4), since only the low-temperature case
$T\ll E_c$ is relevant to our problem. As it follows from Eq. (3.5), the
corresponding 2D component of phonon momentum for small
$q_{\mbox{\scriptsize ex}}$ must also
be small, i.e., $ql_B \ll 1$. Exploiting the commutation relations
(2.9)--(2.10) as well as the properties (2.8) of the ground state, one can
easily find the matrix element $\langle...\rangle$ in Eq. (3.4) for the
operator (2.14):
$$
  \langle...\rangle={\cal N}\langle 0|
  Q_{{\bf q}_{\mbox{\scriptsize ex}}+{\bf q}}
  (A_{\bf q}^++B_{\bf q}^+)Q_{{\bf q}_{\mbox{\scriptsize ex}}}^+|0\rangle
  =2i\sin{\left(\frac{1}{2}l_B^2q
  q_{\mbox{\scriptsize ex}}\sin{\beta}\right)}
  \approx
  il_B^2qq_{\mbox{\scriptsize ex}}\sin{\beta}\,.
                                                    \eqno (3.6)
$$
Finally, substituting
$Mq^{-1}l_B^{-2}\delta(q_{\mbox{\scriptsize ex}}\cos\beta-R_{s,{\bf k}})$
for $\delta(\epsilon_i-\epsilon_f)$ into (3.2), and replacing
summation by integration, we find with Eqs. (3.3),
(3.4) and (3.6) that
$$
  \frac{1}{\tau_s(\mbox{\boldmath $k$})}=\sqrt{\frac{\pi}{2}}\frac{\hbar c_s
  |\gamma|^2M^{5/2}T^{3/2}}{p_0^3l_BL_z
  \tau_{A,s}(\mbox{\boldmath $k$})}kq
  \left[\phi_{3/2}(w)-\phi_{3/2}(w')\right]\,.     \eqno (3.7)
$$
Here
$$
  w=\exp{\left[(\mu-\delta)/T-\left(\hbar c_sMk/l_Bq-ql_B/2\right)^2/
  2MT\right]}\,,\quad
  w'=w\exp{(-\hbar c_sk/T)}\,,                  \eqno (3.8)
$$
and $\phi_{3/2}(w)=w\Phi(w,3/2,1)$, where $\Phi$ is the Lerch
function\cite{gr94}, i.e.,
$$
  \phi_{\nu}(w)=\sum_{\kappa=1}^{\infty}{w^\kappa}/{\kappa^{\nu}}\,.
                                                    \eqno (3.9)
$$

Generally speaking, the inverse time (3.7) is a rather complicated function
of the direction of \mbox{\boldmath $k$} relative to the crystal axes.
Let us set $\mbox{\boldmath
$q$}=(k\sin{\theta}\cos{\varphi},\,k\sin{\theta}\sin{\varphi})$ and
consider the important special cases.

$\;$
\centerline{\bf A. The long-wavelength limit and sound attenuation}

$\;$

To find the 2D absorption of a macroscopic sound wave we consider the case
$$
  kl_B\ll (8MT)^{1/2} \qquad \mbox{and} \qquad T \sim 1\,\mbox{K}\,.
                                                          \eqno (3.10)
$$
Then, excluding a narrow region of $\sin^2{\theta} \lesssim (\hbar c_s
l_B^{-1})^2M/2T,\,$ $w$ does not depend on  \mbox{\boldmath $k$}. Besides, in
case of a completely equilibrium 2DEG
when we neglect the heating, we can set $\mu=0$. Therefore
$$
  w(\theta)=\exp{\left(-\frac{\delta}{T}-\frac{\eta_s}
  {\sin^2{\theta}}\right)}\,,\quad \mbox{where}\quad \eta_s=\frac{\hbar^2
  c_s^2M}{2Tl_B^2}\,. \eqno (3.11)
$$
Substituting (A.1) and (A.2) (see Appendix~I) into (3.7) we
find the acoustic wave attenuation coefficient $\Gamma_s=1/c_s\tau_s$.
Using the condition (3.10) and taking into account that in our case
$\gamma \approx 1$ holds, we obtain the result for different polarizations
$$
  \Gamma_s\sim \frac{10^{-9}}{L_z}kB^{-3/4}T^{1/2}\exp{(-|g\mu B|/T)}
  [{P}_s(\theta)p_s(\varphi)+{\cal D}_s(\theta)]\;
  \mbox{cm}\cdot\mbox{K}^{-1/2}\cdot(\mbox{Tesla})^{3/4}. \eqno (3.12)
$$
For a longitudinal wave the PA interaction leads to the functions
$$
  {P}_l=2\sin^5{\theta}\cos^2{\theta}e^{\delta/T}\phi_{1/2}(w(\theta))
  \quad
  \mbox{and}\quad p_l(\varphi)=\sin^2{2\varphi}\,,             \eqno (3.13)
$$
and the DA interaction gives
$$
  {\cal D}_l=\frac{1.2\cdot10^{-11}k^2}{B}\sin{\theta}
  e^{\delta/T}\phi_{1/2}(w(\theta))\;\mbox{cm}^2\cdot\mbox{T}\,.
                                                   \eqno (3.14)
$$

It is seen that the indicated long-wave condition (3.10)
enables us to take into account the DA interaction only for special
directions of \mbox{\boldmath $k$}, namely, when the term
${P}_l(\theta)p_l(\varphi)$ vanishes.

For transverse polarization we have ${\cal D}_t\equiv 0$. If the wave is
polarized in the direction perpendicular to ${\hat z}$ and \mbox{\boldmath $k$}
(i.e., $t_y=0,\quad t_x=1$ in (A.2)), then we have
${P}_t={P}_{\perp}$, where
$$
  {P}_{\perp}=4\sin^3{\theta}\cos^2{\theta}e^{\delta/T}\phi_{1/2}
  (w(\theta))\,,
  \quad
  \quad p_{\perp}(\varphi)=\cos^2{2\varphi}\,.
$$
For a transverse wave polarized in the plane of the vectors
\mbox{\boldmath
$B$} and \mbox{\boldmath $k$} ($t_x=0\quad t_y=1$) we get
${P}_t={P}_{\parallel}$, where
$$
  {P}_{\parallel}=\sin^3{\theta}(2\cos^2{\theta}-\sin^2{\theta})^2
  e^{\delta/T}\phi_{1/2}(w(\theta))
  \,,\quad\mbox{and}
  \quad p_{\parallel}(\varphi)=\sin^2{2\varphi}\,.
$$
The dependence ${P}_s(\theta)$ is illustrated graphically in Fig. 1 for
the case $T=1\;$K.

Thus, our result is that the bulk sound--wave--attenuation
coefficient, being determined by the PA interaction, is proportional
to $k$ and, naturally, inversely proportional to the
dimension $L_z$. The latter dependence arises from the normalization of
the wave displacement field to the whole sample volume $L^2L_z$ (cf. also
Ref. \onlinecite{io89}). Further, it is easy to estimate how our results
are modified in the case of a surface acoustic wave. The essential
difference is that the displacement field for a surface wave has to be
normalized
to $L^2/|k_z|$, where $k_z$ takes an imaginary value characterizing the
spatial surface wave attenuation in the ${\hat z}$ direction. Actually we
have $|k_z|=bk$,
where $b\sim 1$ is a numerical factor (see Refs. \onlinecite{kn96,lel}).
As a result the
attenuation coefficient is obtained by multiplying Eq. (3.12) by a factor
of order $kL_z$. The corresponding estimate for $B=10\,$T
yields
$$
  \Gamma \sim 10^{-10}k^2\left(\frac{T}{1\,\mbox{K}}\right)^{1/2}
  \exp{\left(-\frac{3\,\mbox{K}}{T}\right)}\cdot\mbox{cm}\,. \eqno (3.15)
$$

The surface acoustic wave attenuation in the half-integer QHR has
been considered in Ref. \onlinecite{kn96}, and the attenuation
coefficient was found to be of order $10^{-5}k$ (correspondingly,
the $k$-independent attenuation has been
obtained in this model for the bulk sound wave\cite{io89,io96}). The Fermi
energy was assumed there to be close to the center of the LL.
According to estimates\cite{kn96}, this zero-temperature result
should be valid up
to $T\sim 1\,$K. At the same time, the calculated  value of $\Gamma$ becomes
very small when the Fermi energy deviates substantially from the center to the
LL
edges. One can actually see that our case, which is only applicable to the
upper edge of the LL, yields a result of the same
order as in Ref. \onlinecite{kn96} (for the LL center) even for $k\sim
10^5\,$cm${}^{-1}$, provided that $T\sim 1\,$K.

The hopping transport and absorption of a surface acoustic wave near the
integer QHR was studied by Aleiner and Shklovskii (Ref. \onlinecite{al94}).
These authors take into
consideration the Coulomb $e$-$e$ interaction within the framework of the
$ac$-hopping conductivity theory\cite{po93}. Then they use a simple
equation connecting the conductivity $\sigma_{xx}(\omega,k)$ due to the
piezoelectric coupling with the attenuation $\Gamma$ for the surface wave.
In the long-wave region $kl_B<0.2$, provided the relationship
$\sigma_{xx}=\varepsilon\xi\omega/6$ holds ($\xi$ is the localization
radius, $\xi \sim l_B$ for $\nu$ close to an integer), one can see
from the results of Ref. \onlinecite{al94} that $\Gamma \sim
10^{-10}k^2\,\mbox{cm}$
for $B=10\,$T and for temperatures $T<\hbar\omega_s$. This is in agreement
with our result if $T\sim 1\,$K.
At the same time our theory gives the dependence $B^{-3/4}\exp{(-|g\mu B|/T)}$
for $\Gamma$,
in contrast to $B^{-1/2}$ in Ref. \onlinecite{al94}, and besides
our attenuation vanishes as $T^{1/2}e^{-\delta/T}$ as $T$
decreases
(of course, as long as the inequality (3.10) is valid), whereas the
corresponding result in Ref. \onlinecite{al94} is temperature independent.

$\;$

\centerline{\bf B. The short-wave limit and the heat absorption}

$\;$

In order to gain insight into such quantities as the rate of ballistic phonon 
absorption or the contribution of the 2DEG to the thermal conduction, we need
to consider the short-wave-length limit.
For $k^2l_B^2> -8MT\ln{(8MT)}$ one can see with the help of
Eqs. (3.7)--(3.9) and (A.1) that the PA absorption for the
longitudinal phonon even at its maximum becomes less than the DA
absorption. Hence in the case which will be
considered now,
$$
   kl_B \gapprox 1\qquad (k\gapprox 10^{6}\,\mbox{cm}^{-1})\,,
   \eqno (3.16)
$$
so that we may neglect the piezoelectric part of the inverse PLT for the $LA$-phonons.
Also in this case the parameter $w$ (3.8) has a narrow maximum at
$\theta=\theta_0$, where
$$
  \sin^2{\theta_0}= 2\hbar c_sM/kl_B^2 \ll 1.            \eqno (3.17)
$$
Thus, only the longitudinal
phonons with \mbox{\boldmath $k$} almost parallel to ${\hat z}$
($q\lapprox 0.1k_z$) effectively interact with the 2DEG. As an example,
in Fig. 2
the angular dependence of the evolution of the PLT (3.7) is shown for a range of
$k$ (the case $s=l$ is considered). Obviously, for the
phonon momentum strictly parallel to ${\hat z}$
(i.e., when $q=0$) Eq. (3.7) also yields zero.
The width of this absorption region is obtained as
$\Delta(\sin^2{\theta}) \sim (\hbar c_sT)^{1/2}
M/k^{3/2}l_B^2$.

Now let us find the rate (or flux) of phonon absorption
${\cal R}_s$ and the heat dissipation rate ${\cal Q}_s$ in a 2DEG:
$$
  {{\cal R}_s\choose {\cal Q}_s}
  =N_s\int n_{\mbox{\scriptsize ph}}^{(s)}(q,k_z){1\choose\hbar
  \omega_{s,{\bf k}}}
  \tau_s^{-1}(\mbox{\boldmath
  $k$})d^3\mbox{\boldmath $k$}\cdot\frac{L^2L_z}{(2\pi)^3}\,. \eqno (3.18)
$$
Here $N_s$ is the spatial density of the $s$-type nonequilibrium
phonons in the sample, and $n^{(s)}_{\mbox{\scriptsize ph}}$ is their
normalized distribution function
($\int n_{\mbox{\scriptsize ph}}^{(s)}
{d^3\mbox{\boldmath $k$}}/(2\pi)^3\equiv 1$).
We assume the phonon distribution to be a broad smooth function in the
$({\hat x},{\hat y})$ plane; therefore
due to the conditions (3.16)--(3.17) we may set
$n_{\mbox{\scriptsize ph}}^{(l)}(q,k_z)\approx
n_{\mbox{\scriptsize ph}}^{(l)}(0,k_z)$ in the integrals (3.18).
Note that the condition (3.16) partly determines the apparent
choice of the distribution $n_{\mbox{\scriptsize ph}}^{(l)}$: this
distribution has to be mainly concentrated
in the range of $k$ satisfying Eq. (3.16). Substituting Eq. (3.7) into
Eq. (3.18) and taking into account that $k\approx k_z$ we find for the
longitudinal polarization:
$$
  {{\cal R}_l\choose {\cal Q}_l}=\frac{N_lL^2M^4
  (\hbar c_l)^2T^2}
  {\pi\tau_Dp_0^3l_B^4}\int_{-\infty}^{+\infty}dk_z{1\choose \hbar
   c_l|k_z|}|k_z
  \gamma(k_z)|^2n_{\mbox{\scriptsize ph}}^{(l)}(0,k_z)\,
  {}\qquad{}\qquad{}\quad{}
$$
$$
  {}\qquad{}\qquad{}\qquad{}\cdot\,\sum_{\kappa=1}^{\infty}
  \frac{\exp{[\kappa(\mu-\delta)/T]}}{\kappa^{2}}
  \Omega_l(\kappa\hbar c_l|k_z|/2T)\,,
                                             \eqno (3.19)
$$
where $\Omega_l(\xi)=(1+\xi)(1-e^{-2\xi})/2\xi\,$.

In contrast, for transverse phonons, when only the PA interaction
determines the absorption, small $k$ of order $(MT)^{1/2}/l_B$
still play the main role.
When the distribution $n^{(t)}_{\mbox{\scriptsize ph}}(\mbox{\boldmath $k$})$ is
sufficiently long-range and provides the same probability for
both transverse polarizations, we may assume
$n^{(t)}_{\mbox{\scriptsize ph}}(\mbox{\boldmath $k$})=n_{\mbox{\scriptsize ph}}^{(t)}(0)$ in Eq. (3.18)
(of course, setting
$\gamma(k_z)\approx\gamma(0)\equiv 1$) and then use
Eq. (A.3). Nevertheless, real
distributions arising from the metal film heaters are really
the Planck ones for small phonon momenta\cite{go96,ak99}, and
$n^{(s)}_{\mbox{\scriptsize ph}}(0)$ goes to infinity. For this reason we assume
that the following conditions take place for the function
$n_{\mbox{\scriptsize ph}}^{(s)}(\mbox{\boldmath $k$})$ characterized by the effective
temperature ${\cal T}$ and by some angle distribution $\Phi(\theta)$
isotropic in the $({\hat x}, {\hat y})$ plane:
$$
\displaystyle{
\begin{array}{l}
  n_{\mbox{\scriptsize ph}}^{(s)}(\mbox{\boldmath $k$})\approx
  \left(\hbar c_s/{\cal T}\right)^2\cdot \Phi(\theta)/k,\quad
  \mbox{when}\quad \hbar c_sk\ll {\cal T},\\
   n_{\mbox{\scriptsize ph}}^{(s)} (\mbox{\boldmath $k$})\ll
  \left(\hbar c_s/{\cal T}\right)^3,\quad\mbox{when}
  \quad \hbar c_sk\gg {\cal T}\,.
\end{array}
} \eqno (3.20)
$$
Then assuming that $\Phi(\theta)\lapprox 1$, and
$$
  {\cal T}\gg T\,,                                \eqno (3.21)
$$
we obtain for $TA$-phonons from Eq. (3.18) in the lowest-order
approximation:
$$
  {{\cal R}_t\choose {\cal Q}_t}=
  \frac{5L^2N_t(M\hbar c_t)^4T^2}{8\alpha\pi p_0l_B^4\tau_P
  {\cal T}^2}
  \sum_{\kappa=1}^{\infty}\frac{\exp{[\kappa(\mu-\delta)/T]}}{\kappa^2}\,
  {}\qquad{}\qquad{}\qquad{}\qquad{}
$$
$$
  {}\qquad{}\qquad{}\qquad{}\qquad\cdot\int_0^{\pi}d\theta
  \Phi(\theta)(9\cos^4{\theta}-\cos{2\theta})
  {\sin{\theta}
        \choose 2^{5/2}\hbar c_tl_B^{-1}\sqrt{\pi
        MT/\kappa}}\,,
                                                   \eqno (3.22)
$$
where $\alpha=c_t^2/c_l^2$, which is $\approx 0.36$ for GaAs.
We have substituted into Eq. (3.18) the formula (3.7) for $\tau_t^{-1}$
employing Eq. (A.3) from Appendix~I.

Dividing the value ${\cal R}_s$ by $N_sL^2L_z$, one 
finds for the appropriate effective inverse PLT,
$$
  {\left[{{\tau^{(l)}}_{\mbox{\scriptsize eff}}}^{-1}\atop
  {{\tau^{(t)}}_{\mbox{\scriptsize eff}}}^{-1}\right]}
  \sim \frac{M^4(\hbar c_l)^2T^2}{L_zp_0l_B^4}
  \cdot\exp{[(\mu-\delta)/T]}\;
  \cdot
  {\left[(\lambda/p_0)^{2}\tau_D^{-1}
   \atop (\hbar c_t/{\cal T})^{2}\tau_P^{-1}
    \right]}\,,                                             \eqno (3.23)
$$
and the ratio ${\tau^{(l)}}_{\mbox{\scriptsize eff}}/
{\tau^{(t)}}_{\mbox{\scriptsize eff}}\sim
  (\hbar c_t p_0/{\lambda\cal
  T})^2\tau_D/\tau_P$,
where $\lambda=|\gamma(l_B{\cal T}/\hbar c_l)|$. This means that the
absorption of longitudinal phonons is larger by a factor of order
$100$ than that of transverse phonons for the same
distribution (3.20)--(3.21). (It is also assumed that
${\cal T}\sim 10\,$K and $\lambda \sim 1$). Analogous
results are obtained for the ratio ${\cal Q}_t/{\cal Q}_l$.

Finally, in order to find the absolute magnitudes of the relevant absorption
characteristics we should determine the SW chemical potential $\mu$.
This is found from the conservation of the total 2DEG spin $S$ or, what is the 
same, the equation for the conservation of the total number of SWs:
$$
  N_{\mbox{\scriptsize SW}}={\cal N}-S=-MT{\cal N}\ln{\left\{1-\exp{[(\mu-\delta)/T]}\right\}}\,.
                                                        \eqno (3.24)
$$
This is simply the number of free 2D Bose particles
at temperature $T$ and chemical potential $\mu$ (see, e.g.,
Ref. \onlinecite{di96}). Equating this value to the same one at the initial
temperature $T_0$ and at zero chemical potential (describing the 2DEG
before the heating was started) we get
$$
  \exp{[(\mu-\delta)/T]}=1-\left[1-\exp{(-\delta/T_0)}\right]^{T_0/T}\,.
                                                      \eqno (3.25)
$$
Thus, $\mu$ is determined by
temperatures $T$ and $T_0$, and to first order in
$\Delta T/T_0$ we may substitute
$\exp{(-\delta/T_0)}$ for $\exp{[(\mu-\delta)/T]}$ into Eqs. (3.19),
(3.22), and (3.23).

Specifically for the times (3.23) in the case
of a field $B=10\,$T one can estimate
$$
  \frac{1} {\tau^{(s)}_{\mbox{\scriptsize eff}}}\sim f_s\frac{T^2_0}{L_z}
  \exp{\left(-\frac{3\,\mbox{K}}{T_0}\right)}\cdot
  \frac{\mbox{cm}}{\mbox{K}^2\cdot\mbox{s}}\,,         \eqno (3.26)
$$
where $f_l\sim 1\div 10^{-1}$  for $LA$ phonons, and  $f_t\sim 10^{-1}\div
10^{-2}$ for $TA$ modes.

The quantities (3.23) and (3.26) determine the 2DEG contribution to
the inverse total thermal conductance which may be estimated by
means of the kinetic formula
$$
  \Delta(\Xi ^{-1})\sim C_{\mbox{\scriptsize ph}}^{-1}
  \left(\frac{1}{c_l^2\tau_{\mbox{\scriptsize eff}}^{(l)}}
  +\frac{1}{c_t^2\tau_{\mbox{\scriptsize eff}}^{(t)}}\right),
                                                         \eqno (3.27)
$$
where $C_{\mbox{\scriptsize ph}}$ is the 3D lattice heat capacity.
One can see that even
under favourable experimental conditions we have
$\Xi\Delta(\Xi ^{-1})\lesssim 10^{-4}$. Therefore, a small value of
Eq. (3.27) does not permit us to consider our
mechanism relevant to heat absorption under the experimental conditions of
Ref. \onlinecite{ei87}, where the sensitivity allowed
only the variations $\Xi\Delta(\Xi ^{-1})>5\cdot 10^{-3}$ to be measured.

\section{Phonon absorption at zero temperature and spin state change
 (the second absorption channel)}
If the 2DEG temperature goes to zero then the above results of the first
absorption channel vanish. On the other hand the SO terms in Eq. (2.14)
can give a substantial contribution to the inverse PLT even
at zero $T$. These terms allow the absorbed phonon to create a
spin wave, thereby changing the spin state. Evidently this
is the transition between the 2DEG states $|i\rangle=|0\rangle$ and
$\langle f|=\langle\mbox{\boldmath $q$}|$, provided the absorbed
phonon has the wave vector $\mbox{\boldmath $k$}=(\mbox{\boldmath
$q$},k_z)$. Only phonons with energies higher than the threshold
$\delta$ can be absorbed. The quantity
$$
  {\tilde W}_s(\mbox{\boldmath $q$}_{\mbox{\scriptsize ex}},
  \mbox{\boldmath $k$})=
  \frac{2\pi}{\hbar}
  \left|\tilde{\cal M}_s(\mbox{\boldmath
  $q$})\right|^2\delta[\epsilon(q_{\mbox{\scriptsize ex}})-
  \hbar\omega_{s,{\bf k}}]
  \delta_{{\bf q}_{\mbox{\scriptsize ex}},{\bf q}}       \eqno(4.1)
$$
is the probability of this process, and the kinematic
relation holds:
$
  \hbar c_s\sqrt{k_z^2+q^2}=\delta +(ql_B)^2/2M.
$
Therefore
$$
  k_z=\pm K(q)\approx\pm \left(\delta+q^2l_B^2/2M\right)/
  \hbar c_s\,.                                     \eqno (4.2)
$$
One can see that only a selected resonant group of phonons takes part in this 
process. The possible magnitude of $k_z$ (4.2) always satisfies the condition
$|k_z|\gg q$ as well as $k_zl_B \gapprox 1$ for our QHR parameter region. 
In addition, just as in the short-wave limit of the first absorption channel 
(see the previous section), we again find that only the phonons with momenta 
almost parallel to the normal ${\hat z}$ interact effectively with the 2DEG.

The calculation of the matrix element ${\cal M}_s$ of the Hamiltonian (2.11)
reduces to the calculation of
$\langle\mbox{\boldmath $q$}|H_{\mbox{\scriptsize e,ph}}|0\rangle$ and
is substantially simplified by the commutation relation (2.10) and
Eqs. (2.8). Eventually Eqs. (2.8)--(2.14) enable us to obtain
$$
  |\tilde{\cal M}_s|^2=
  \frac{\hbar^2c_sk|\gamma(k_z)|^2
  [(u^2+v^2)q^2-2uvq_xq_y]}{4L_zp_0^3
  \tau_{A,s}(\mbox{\boldmath $k$})}\,.                 \eqno (4.3)
$$
From this the inverse life-time of a nonequilibrium phonon with the
mechanism of SW creation,
$$
  \frac{1}{{\tilde \tau}_s(\mbox{\boldmath $k$})}=
  \sum_{{\bf q}_{\mbox{\scriptsize ex}}}
  {\tilde W}_s(\mbox{\boldmath $q$}_{\mbox{\scriptsize ex}},
  \mbox{\boldmath $k$})=
  \frac{|\tilde{\cal M}_s|^2L_zk}{\hbar^2c_sk_z}
  \left[\delta_{k_z,K(q)}+\delta_{k_z,-K(q)}\right], \eqno (4.4)
$$
is readily obtained.
After averaging over the \mbox{\boldmath $q$} directions and also over
the \mbox{\boldmath $t$} directions for $TA$ phonons (see the Appendix~I)
we have
$$
  \frac{1}{{\tilde \tau}^{(s)}(q,k_z)}=\int_0^{2\pi}\frac{d\varphi}{2\pi}
  \frac{1}{{\tilde \tau}_s(\mbox{\boldmath $k$})}\approx
  \frac{|\gamma|^2(u^2+v^2)q^2K(q)}
  {4p_0^3\tau_{0s}(q,k_z)}
  \left[\delta_{k_z,K(q)}+\delta_{k_z,-K(q)}\right]\,, \eqno (4.5)
$$
 where
$$
  1/\tau_{0l}\approx 1/\tau_D,\quad{}\qquad 1/\tau_{0t} \approx
  \frac{5p_0^2q^2}{\alpha \tau_Pk^4}\,.           \eqno (4.6)
$$
The expression for $1/\tau_{0t}$ follows from Eq. (A.3) of
Appendix~I; numerical calculation gives
$\tau_{0t}\approx 40k^4/q^2\,$ps${}\cdot$nm${}^2$.

In spite of the small factor $u^2+v^2$ for such
resonant phonons, the inverse time in (4.5) is comparatively large
(of order 10${}^{6}\,$s${}^{-1}$
for $q\sim 0.1\,$nm${}^{-1}$) and
does not depend on the temperature. The reason for this lies in the fact 
that the rate of SW creation and of phonon absorption
is proportional to the large degeneracy factor ${\cal N}$ of the LL,
whereas the corresponding value of the inverse PLT (3.7)
calculated in the previous Section is
proportional only to the SW density which is exponentially
low (at low $T$) due to Eqs. (3.1) and (3.3).

The effective inverse PLTs are more important for the applications.
These quantities, which characterize the rate of SW creation
$\tilde{\cal R}_s$ (equivalent to the phonon absorption rate) and of
heat absorption $\tilde{\cal Q}_s$ are determined as follows:
$$
  \frac{1}{{\tilde
  \tau}^{(s)}_{\mbox{\scriptsize eff}}}=
  \int\frac{n_{\mbox{\scriptsize ph}}^{(s)}(q,k_z)}
  {{\tilde \tau}^{(s)}(q,k_z)}
  \frac{d^3\mbox{\boldmath $k$}}{(2\pi)^3}\,,\qquad
  \frac{1}{{\tilde \tau}^{(s)}_{Q}}=\frac{\hbar c_s}{\delta}\int
  \frac{n_{\mbox{\scriptsize ph}}^{(s)}(q,k_z)}
  {{\tilde \tau}^{(s)}(q,k_z)}
  \frac{kd^3\mbox{\boldmath $k$}}{(2\pi)^3}\,.\eqno (4.7)
$$
Accordingly, from Eq. (3.18) we have
$$
  \tilde{\cal R}_s=N_sL^2L_z/{\tilde
  \tau}^{(s)}_{\mbox{\scriptsize eff}}\,,\quad\mbox{and}
  \quad
  \tilde{\cal Q}_s=N_sL^2L_z\delta/{\tilde \tau}^{(s)}_{Q}\,.\eqno (4.8)
$$
As a result, using Eqs. (4.5)-(4.6) we obtain
$$
  {\left[1/{{\tilde \tau}_{\mbox{\scriptsize eff}}}^{(l)}\atop
   1/{{\tilde \tau}^{(l)}}_{Q}\right]}= \frac{(u^2+v^2)(\hbar c_lM)^2}
  {4\pi p_0^3l_B^4L_z\tau_D}\int_{|k_z|>k_{0l}}n^{(l)}_{\mbox{\scriptsize ph}}(0,k_z)
  |\gamma(k_z)|^2
  (|k_z|-k_{0l})
  {\left[|k_z|\atop \hbar c_lk_z^2/\delta\right]}dk_z\,, \eqno (4.9)
$$
and
$$
{\left[1/{{\tilde
\tau}^{(t)}}_{\mbox{\scriptsize eff}}\atop
1/{{\tilde \tau}^{(t)}}_{Q}\right]}
 =\frac{5(u^2+v^2)(\hbar c_tM)^3}{2\pi\alpha
  p_0l_B^6L_z\tau_P}\int_{|k_z|>k_{0t}}n^{(t)}_{\mbox{\scriptsize ph}}(0,k_z)|\gamma(k_z)|^2
  (|k_z|-k_{0t})^2
  {\left[|k_z|^{-3}\atop \hbar c_t/k_z^2\delta\right]}dk_z
                         \,,
$$

${}\;{}$

\noindent
where $k_{0l}=\delta/\hbar c_l$ and $k_{0t}=\delta/\hbar c_t$. Here we
have assumed that for small $q$ [that is for $q \sim (M\hbar
c_s)^{1/2}/l_B^{3/2}$, which give the
main contribution to the integrals in (4.7)] 
$n^{(s)}_{\mbox{\scriptsize ph}}(q,k_z)\approx 
n^{(s)}_{\mbox{\scriptsize ph}}(0,k_z)$.

We now compare the values found here with
the analogous ones of the first absorption channel (3.23). One can
estimate for the distribution function (3.20)  the ratio of
PLTs for LA phonons:
$$
  \left[{\tilde \tau}^{(l)}_{\mbox{\scriptsize eff}}\right]^{-1}/
  \left[{\tau}^{(l)}_{\mbox{\scriptsize eff}}\right]^{-1}\sim
  \frac{(u^2+v^2)\exp{[|g\mu_bB|/T_0]}}{8\pi (MT_0)^2}\,,   \eqno (4.10)
$$
and for the TA mode:
$$
  \left[{\tilde \tau}^{(t)}_{\mbox{\scriptsize eff}}\right]^{-1}/
  \left[{\tau}^{(t)}_{\mbox{\scriptsize eff}}\right]^{-1}\sim
  \frac{(u^2+v^2)\hbar^2 c_t^2\exp{[|g\mu_bB|/T_0]}}
  {MT_0^2l_B^2|g\mu_bB|}\,.
                                                             \eqno (4.11)
$$
Here we have assumed ${\cal T}\gg T_0$, and $|T-T_0|\ll T_0$.
Substitution of the characteristic numerical magnitudes for the
quantities entering in Eqs. (4.10)--(4.11) results in the observation that:
in spite of the small spin-orbit parameter $u^2+v^2$, the inverse PLT for
the second channel at $T_0\lapprox 1\,$K may be
comparable or even larger than that corresponding to the first channel.

\section{Quasi-equilibrium temperature and spin momentum of 2DEG
in the presence of nonequilibrium phonons}

So far we have calculated the absorption rates (in the form of
the phonon-number and heat absorption fluxes) determined exclusively
by the nonequilibrium phonons. However, one should bear in mind
that these calculations leave the 2DEG temperature
undetermined. Below we study the growth of $T$ due to the processes 
considered above, since in a real
experiment the observation time may be of the order of or even much
longer than  $\tau_{\mbox{\scriptsize eff}}$ and
${\tilde \tau}_{\mbox{\scriptsize eff}}$ found in the previous
sections. Therefore, it is of interest to find the quasi-equilibrium $T$
and $\mu$ for the SW gas in the presence of permanent phonon
pumping. Here in addition to finding these, we will
estimate the amount of time required for the dynamic
equilibrium to be established. Recall that the SW
distribution function in (3.3) is supposed to apply
always, since the time required for establishing
thermal equilibrium among the SWs is relatively short (see Appendix~II). 

The dependence on $t$ of $T(t)$ and $\mu(t)$ is determined by
the following balance equations for the SW number and heat :
$$
  \sum_s\left[\tilde{\cal R}_s-\tilde{\cal R}^{(0)}_s\right]=
  dN_{\mbox{\scriptsize SW}}/dt,
  \qquad
  \sum_s\left[{\cal Q}_s+\tilde{\cal Q}_s-{\cal Q}^{(0)}_s-\tilde{\cal
  Q}^{(0)}_s\right]=dE_{\mbox{\scriptsize SW}}/dt
  \,.
                                                       \eqno (5.1)
$$
The fluxes $\tilde{\cal R}_s$, ${\cal Q}_s$ and
$\tilde{\cal Q}_s$ have been found in the previous Sections. The flux
$\tilde{\cal R}^{(0)}_s$ is the rate of spin relaxation to
its equilibrium magnitude at $T_0$. In other words, it
is the rate of SW annihilation (which is the process inverse to that 
corresponding to the second channel) due to acoustic phonon emission.
The heat fluxes ${\cal Q}^{(0)}_s$
and $\tilde{\cal Q}^{(0)}_s$ (correspondingly of the first and
the second channel) are
the back flows carrying the heat from
the overheated 2DEG to the lattice held at a fixed temperature
$T_0$. (The overheating $\Delta T=T-T_0$, which occurs due to
the presence of nonequilibrium phonons, causes these
fluxes from the SW gas to the equilibrium phonon bath at $T_0$).
The SW number $N_{\mbox{\scriptsize SW}}$ is determined by the
formula (3.24), and the quantity
$E_{\mbox{\scriptsize SW}}(T,\mu)$ is the spin--wave--excitation part
of the 2DEG energy, which is determined by
$$
  E_{\mbox{\scriptsize SW}}=
  N_{\mbox{\scriptsize SW}}\delta+{\cal N}T^2M\phi_2(\varrho)\quad
  \left(\mbox{where}\quad\varrho=\exp{[(\mu-\delta)/T]}\right) \eqno (5.2)
$$
for 2D Bose particles with the quadratic spectrum (2.4).
By definition, the flux
$\sum_s{\cal Q}^{(0)}_s$ describes for the first absorption channel the
energy exchange with equilibrium phonons without a change in the SW
number; we have
$$
  {\cal Q}^{(0)}_{s}=\hbar\sum_{\bf k}\sum_{i\ne f}\omega_{s,{\bf k}}
  W_{if}(\mbox{\boldmath $k$})
  \left\{b_T(\epsilon_i)[1+b_T(\epsilon_f)]\left[1+n^{(0)}_{T_0}(k)\right]-
  b_T(\epsilon_f)[1+b_T(\epsilon_i)]n^{(0)}_{T_0}(k)\right\}\,.
                                                      \eqno (5.3)
$$
Here $W_{if}$ is determined again by Eq. (3.2) with
the argument of the $\delta$-function replaced by $\epsilon_i-\epsilon_f-
\hbar\omega_{s,{\bf k}}$; $b_T(\epsilon)$ is, as before, the function 
presented in (3.3), and
$n^{(0)}_{T}$ is the Planck function for the equilibrium phonons,
$n^{(0)}_{T}(k)=1/\left[{\exp{(\hbar\omega_{s,{\bf k}}/T)}-1}\right]$.
The right side in Eq. (5.3) may be easily transformed in such a way
that we obtain an expression similar to that in Eq. (3.18) for ${\cal Q}_{s}$
with PLT (3.1). In so doing 
$N_sn_{\mbox{\scriptsize ph}}^{(s)}(k)$ there
should be replaced by $n^{(0)}_T(k)-n^{(0)}_{T_0}(k)$, however
the following manipulations are quite analogous to those
done when deriving the formulae (3.7), (3.19), and (3.22).
Fortunately, the rather
cumbersome expression for the sum $\sum_s{\cal Q}^{(0)}_s$ may be
simplified,
provided that the temperature is reasonably low. Namely, if
$$
  T\lapprox M^{1/5}(\hbar
  c_l/l_B)^{6/5}(p_0l_B)^{4/5} \eqno (5.4)
$$
(specifically $T\lesssim 1\,$K), then the
piezoelectric interaction gives the main contribution to the sum
$\sum_s{\cal Q}^{(0)}_s\approx {\cal Q}^{(0)}_P$, where
$$
  {\cal Q}^{(0)}_P=\frac{45(\hbar c_lL)^2M^{9/2}T^{5/2}(T-T_0)}
  {4\pi^{1/2}p_0l_B^5\tau_P}\exp{[(\mu-\delta)/T]}\,.
                                                   \eqno (5.5)
$$
We assume further that
$$
\exp{[(\mu-\delta)/T]}\ll 1  \,,                 \eqno (5.6)
$$
which has already been used when obtaining expression (5.5).

We determine ${\tilde {\cal R}}_s^{(0)}$ and ${\tilde {\cal
Q}}_s^{(0)}$ in Eq. (5.1) from the expressions
$$
  {\tilde{\cal R}^{(0)}_s\choose \tilde{\cal Q}^{(0)}_s}=
  \sum_{{\bf k}}\sum_{{\bf q}_{\mbox{\scriptsize ex}}}\tilde{
  W}_s(\mbox{\boldmath $q$}_{\mbox{\scriptsize ex}},
  \mbox{\boldmath $k$}){1\choose
  \hbar\omega_{s,{\bf k}}}\left[
  b_T(\mbox{\boldmath $q$}_{\mbox{\scriptsize ex}})-n^{(0)}_{T_0}
  (\mbox{\boldmath $k$})\right]\,.                    \eqno (5.7)
$$
The sum $\sum_s {\tilde {\cal R}}_s^{(0)}$ in Eq. (5.1) has been
calculated in Ref. \onlinecite{di96} for the case $T=T_0$ (c.f. Eq.
(6.27) herein) and can be determined in a similar way in
our case. Assuming that $T \gg T_0$ and taking into
account the condition in (5.6), we neglect the PA
interaction
and obtain $\sum_s{\tilde{\cal R}^{(0)}_s}\approx
\tilde{\cal R}^{(0)}_D,\quad \sum_s\tilde{\cal Q}^{(0)}_s\approx
\tilde{\cal R}^{(0)}_D\delta$, where
$$
  \tilde{\cal R}^{(0)}_D=
  \frac{(u^2+v^2)(LMT)^2\delta}
  {2\pi \hbar c_lp_0^3l_B^4\tau_D}\exp{[(\mu-\delta)/T]}\,.\eqno (5.8)
$$
With all terms on the left sides of Eqs. (5.1) thus determined, we can 
find the dependence on $t$ of $T(t)$ and $\mu(t)$. However, this would be
meaningful only for
comparison with a certain experiment. For the present we restrict
ourselves to consideration of two special cases. 
Assuming further that only LA phonons are pumped into the sample,
Eqs. (5.1) transform into
$$
  \tilde{\cal R}_l-\tilde{\cal
  R}_D^{(0)}=dN_{\mbox{\scriptsize SW}}/dt,\quad\mbox{and}\quad
  {\cal Q}_l + \tilde{{\cal Q}_l}-{\cal Q}_P^{(0)}-\tilde{\cal
  R}^{(0)}_D\delta=dE_{\mbox{\scriptsize SW}}/dt\,.
                                              \eqno  (5.9)
$$

$\;$

\centerline{\bf A. Appreciable initial temperature;
predominance}
\centerline{\bf of the first absorption channel.}

$\;$
Here we assume that the initial temperature satisfies the conditions 
$\Delta T = T - T_0 \ll T_0 < \delta$. We further assume that Eqs. (5.4) 
and (5.6) apply. We observe that under these conditions, the terms 
corresponding to the second absorption channel are much smaller than the 
others in the first equation in Eq. (5.9). Supposing again that the
features of the nonequilibrium phonon distribution expressed by Eqs.
(3.20) hold, we find from
the equation ${\cal Q}_l={\cal Q}^{(0)}_P$ (see Eqs. (3.19) and (5.5))
the temperature shift
$$
 \Delta T \sim
 \frac{N_ll_B{\cal T}\tau_P}{p_0^2(MT_0)^{1/2}\tau_D}\,.
                                                   \eqno (5.10)
$$
For $N_l = 10^{15}\,$cm${}^{-3}$,
${\cal T}=10\,$K, $l_B=8\,$nm and $(MT_0)^{1/2}=0.1$ we have
$\Delta T\sim 100\,$mK. The resulting overheat (5.10) is
determined only by the first absorption channel. Hence one can
ignore the SO channel of absorption only for not too low initial
temperatures $T_0>0.1\,$K.

Let us now also estimate the time $(\Delta t)_T$ needed to establish the
quasi-equilibrium temperature. For $t\sim (\Delta t)_T$
the three terms
${\cal Q}_P^{(0)}$, ${\cal Q}_l$ and
$dE_{\mbox{\scriptsize SW}}/dt$ in Eq. (5.1)
become of the same order. We equate the expression (5.5) and the
rate of heating
$dE_{\mbox{\scriptsize SW}}/dt\approx C_N\Delta T/\Delta t$, where
$C_N$ is
the heat capacity of the 2D Bose gas at constant SW number (3.24) (the
inequality (5.6)
enables us to find $C_N\approx N_{\mbox{\scriptsize SW}}$). The result is
$$
 (\Delta t)_T \sim \frac{p_0l_B^3\tau_P}{45(\hbar c_l)^2M^{7/2}T^{3/2}}
  \sim 10^{-7}\left(\frac{B}{1\;\mbox{Tesla}}\right)^{1/4}
  \left(\frac{T}{1\,\mbox{K}}\right)^{-3/2}\, \mbox{s}\,.
                                                   \eqno (5.11)
$$
Note that this value does not depend on the level of phonon pumping
$N_l$. The time $(\Delta t)_T$ is found to be shorter than the
spin relaxation time\cite{di96} (see the expression for $(\Delta
{\tilde t})_T$ in Eq. (5.18) below)
down to temperatures $T_0\sim 10\,$mK, i.e., leaving the principle
of SW number conservation intact.

$\;$

\centerline{\bf B. Negligible initial temperature.}

$\;$

Now let us study the opposite case as that considered
above, where the
initial 2DEG temperature is assumed to be very low,
$T_0\ll T$, and find the 2DEG final temperature $T$ and the chemical
potential $\mu$. To this end we set the terms on the right-hand sides
of Eqs. (5.9) equal to zero, substitute  Eqs. (3.19),
(4.8), (4.9), (5.5), and (5.8) for the fluxes in Eqs. (5.9),
take $T_0=0$ and employ the conditions (3.20) of phonon
distribution. Upon this and some algebraic
manipulations, we obtain the following results:
$$
  \exp{[(\mu-\delta)/T]}=\frac{N_l(\hbar c_l)^3}{2T^2\delta}
  \int_{|k_z|>k_{0l}}n^{(l)}_{\mbox{\scriptsize ph}}(0,k_z)|\gamma(k_z)|^2
   (|k_z|-k_{0l})|k_z|dk_z\,,                             \eqno (5.12)
$$
and
$$
  T^{3/2}={T}_1^{3/2}+{\tilde T}^{3/2}\,,            \eqno (5.13)
$$
where
$$
  {T}_1^{3/2}=\frac{2N_l\hbar c_l(\tau_P/\tau_D)}
  {45(\pi M)^{1/2}p_0^2}\int_{-\infty}^{+\infty}dk_z
  n_{\mbox{\scriptsize ph}}^{(l)}(0,k_z)|\gamma(k_z)|^2|k_z|^2,
                                                     \eqno (5.14)
$$
and
$$
  {\tilde T}^{3/2}={\cal C}\frac{(u^2+v^2)(\tau_P/\tau_D)l_B
  {\cal T}_{\mbox{\scriptsize eff}}\delta}
  {M^{5/2}(\hbar c_l)^3 p_0^2}                  \eqno (5.15)
$$
with ${\cal C}=\pi^{7/2}/675\zeta(3)=0.0677$. The quantity
${\cal T}_{\mbox{\scriptsize eff}}$ in the last equation is
$$
  {\cal T}_{\mbox{\scriptsize eff}}=\frac{30\zeta(3)\hbar c_l}{\pi^4}\cdot
  \frac{\int_{|k_z|>k_{0l}}n^{(l)}_{\mbox{\scriptsize ph}}(0,k_z)|\gamma(k_z)|^2
  (|k_z|-k_{0l})^2|k_z|dk_z}{\int_{|k_z|>k_{0l}}n^{(l)}_{\mbox{\scriptsize ph}}(0,k_z)
  |\gamma(k_z)|^2(|k_z|-k_{0l})|k_z|dk_z}\,.      \eqno (5.16)
$$
If the distribution (3.20) is the Planck distribution with 
${\cal T}\gg \hbar c_l/l_B$, it follows that 
${\cal T}_{\mbox{\scriptsize eff}}={\cal T}$.

Thus, the final quasi-equilibrium temperature is determined by
two terms corresponding to
different types of phonon dissipation. The first one in the formula
(5.13) occurs due to the first absorption channel and is proportional
to the level of the phonon excitation, $N_l$. One can see that the
condition (5.6) together with Eq. (5.12) restricts this value to 
$N_l\lapprox 10^{15}\,\mbox{cm}^{-3}$,
which is appropriate for a real experimental situation (see,
e.g., Refs. \onlinecite{ch90,ke92,we86}). For ${\cal T}=10$~K one has the
following result
$$
  T_1\ll {\tilde T}\sim (0.01\div 0.05)\left({\frac{B}
  {1\,\mbox{Tesla}}}\right)^{1/2}\;
  \mbox{K} \sim 20\div 200\mbox{mK}\,. \eqno (5.17)
$$
The basic mechanism of such 2DEG heating, starting from a very low
temperature ($T_0<{\tilde T}$), is related to the
second absorption channel. In this way the final temperature turns out to be
independent of the nonequilibrium phonon density $N_l$ and depends
only on the effective nonequilibrium phonon
temperature (5.15). 

Let us now obtain an estimate for the time $\tilde{(\Delta t)_T}$ required for 
the dynamic equilibrium to establish. Analogously to the calculation
of ${(\Delta t)_T}$, the relationships $\tilde{\cal
Q}_l\sim\tilde{\cal R}^{(0)}_D\delta\sim
dE_{\mbox{\scriptsize SW}}/dt$ hold, provided that 
$t\sim \tilde{(\Delta t)_T}$. Here, according to Eq. (5.2) and the
condition in (5.6), $E_{SW} \approx \delta\cdot M{T}
{\cal N}\exp{[(\mu-\delta)/{T}]}$. Therefore, making use  of Eq.
(5.8) for $T\sim {\tilde T}$ we obtain
$$
  \tilde{(\Delta t)_T}=\frac{\hbar c_ll_B^2p_0^3\tau_D}{(u^2+v^2)M{\tilde
  T}\delta}\sim
  10^{-5}\left(\frac{1\,\mbox{Tesla}}{B}\right)^{1/2}
  \left(\frac{1\,\mbox{K}}{\tilde{T}}\right)\,\mbox{s}\propto
  1/B.                                      \eqno (5.18)
$$
This is the spin relaxation time\cite{di96} for temperature
${\tilde T}$ (5.17).

The details of how the final temperature is established in the case
$T_0\ll {\tilde T}$ are as follows. The generation of the SW is
determined by the second channel of the phonon absorption. The 
resulting SWs which have energies on the order of 
${\cal T}_{\mbox{\scriptsize eff}}$ lose it rapidly (during the time 
interval $(\Delta t)_T$ in (5.11), where one has to substitute
${\cal T}_{\mbox{\scriptsize eff}}$ for $T$), and thus become ``cool'' 
through phonon emission process associated with the first channel.
Provided the ``cooling'' during a short life-time is weak, it follows
that the shorter the life-time, the greater the mean
SW energy ($\sim {\tilde T}$) becomes. This life-time 
$(\Delta {\tilde t})_T$, given in Eq.(5.18), which is
the spin relaxation time\cite{di96}, it is inversely
proportional to $u^2+v^2$, and thus ${\tilde T}$ increases
with the growth of
the SO coupling. Besides, the additional ``warming" of the available
SWs occurs due to the first absorption channel which determines
the value of $T_1$. Naturally, the intensity of the latter effect
becomes larger as the phonon density $N_l$ increases.

In contrast, the SW number and spin change, which in our case
equal to
$$
  N_{\mbox{\scriptsize SW}}=|\Delta S|=
  2S_0MT\exp{[(\mu-\delta)/T]}\qquad(S_0={\cal
  N}/2)\,,
                                              \eqno (5.19)
$$
is according to
Eq. (5.12) proportional to the density $N_l$, so that the spin change 
$\Delta S$ satisfies
$$
  \Delta S/S_0 \sim 10^{-16}B^{-3/2}
  N_{l}\cdot\mbox{cm}^3\cdot(\mbox{Tesla})^{3/2}    \eqno (5.20)
$$
(recall that $\Delta S_z=\Delta S$). If one were able to
create a distribution with a sufficiently large number of the
resonant phonons
($> 5\cdot 10^{15}\,$cm${}^{-3}$), then the observable deviation
of the spin number from the ground state value $S_0$ could be
obtained.

\section{Summary}
The main results of the present work are as follows:

First, there are two different absorption channels in the problem of
acoustic phonon absorption by 2D spin dielectric. The basic
result is the PLT calculation
$$
  1/\tau_{\mbox{\scriptsize ph}}(\mbox{\boldmath $k$})=1/\tau_s(\mbox{\boldmath $k$})+
  1/{\tilde \tau_s}(\mbox{\boldmath $k$})         \eqno (6.1)
$$
(see Eqs. (3.7) and (4.4)). It is a building block in the
study of the effects of sound attenuation and heat absorption, though
the value (6.1) itself cannot be measured directly in the experiments.
The specific averaged time characteristics for the first absorption channel
are presented by Eqs. (3.23) and (3.26).

Second, in spite of the small spin--orbit parameters the temperature 
independent value $1/{\tilde \tau}^{(s)}_{\mbox{\scriptsize eff}}$ (4.7) in 
case of the second
absorption channel may be comparable or even higher than the corresponding
first channel value at $T\lesssim 1\,$K (see Eqs. (4.10)--(4.11)).

Third, according to our calculation, the acoustic bulk and surface wave
absorption by a 2D spin dielectric (Eqs. (3.12) and (3.15)) may be of the
same order or even stronger than the corresponding value in a 2D conductor
(e.g., if filling is $\nu \simeq n+1/2$).

Fourth, even though the 3D sample temperature is negligible ($T_0<0.02\,$K),
the 2DEG temperature due to the phonon heating turns out to be
substantially higher than $T_0$, being independent of the nonequilibrium
phonon density over a wide range:
$10^{11}\,\mbox{cm}^{-3}<N_s<10^{15}\,\mbox{cm}^{-3}$ (see Eqs.
(5.15)--(5.16) and Appendix~II).

Fifth, phonon absorption could lead to an observable change of the
total spin momentum (5.19)--(5.20), if one creates a sufficiently
large number of nonequilibrium phonons in a sample. At the same
time, the evident experimental
difficulty is that one should be able to pump a significant
amount of
nonequilibrium phonons into the sample, keeping the 2DEG
temperature rather low.

And finally, the method of excitonic representation used is straightforward
and very suitable to calculate the relevant transition matrix
elements between the 2DEG states.

${}\;{}$

\centerline{\large\bf Acknowledgements}

${}\;{}$

The author is grateful for the hospitality of
the Max Planck Institute for
Physics of Complex Systems, where the main part of this work was carried
out. Also the author acknowledges the useful information on the background
experiments communicated by V. Dolgopolov, I. Kukushkin,
Y. Levinson, and V. Zhitomirskii. Special thanks are due to
D. Garanin, and V. Zhilin for help in computations.

${}\;{}$

\centerline{{\large\bf Appendix I: the three-dimensional PLTs} \mbox{\large
\boldmath $\tau_{A,s}$}}

${}\;{}$

The derivation of the expressions for $\tau_{A,l}$ and $\tau_{A,t}$ is
analogous to that of similar formulae in Ref. \onlinecite{di96}. The only
difference is that now we consider a more realistic case where $c_l\ne c_t$.
Nevertheless, as in the previous work we again use  the isotropic model
neglecting the dependences of the sound velocities on the orientation with
respect to the crystal axis. This enables us to take into account
the deformation and piezoelectric fields
independently\cite{gale87}, so that the squared value
$\left|U_s(\mbox{\boldmath $k$})\right|^2$ can be transformed
to the sum of the appropriate squares of each type of
interaction. In addition, the transverse phonons in a cubic crystal
do not induce a deformation field.

If we take ${\hat x},\;{\hat y},\;{\hat z}$ to be the
directions of the principal crystal axes, then for a longitudinal phonon
we have
$$
  \frac{1}{\tau_{A,l}(\mbox{\boldmath $k$})}=
  \frac{1}{\tau_D} +
  \frac{45p_0^2}{k^8\tau_P}q_x^2q_y^2k_z^2\,,     \eqno (A.1)
$$
and for a transverse phonon
$$
  \frac{1}{\tau_{A,t}(\mbox{\boldmath $k$})}=
  \frac{5(c_lp_0)^2}{c_t^2 q^2k^8\tau_P}\left[t_x\cdot(q_y^2-q_x^2)kk_z +
  t_y\cdot(2k_z^2-q^2)q_xq_y\right]^2\,.                       \eqno (A.2)
$$
Here $t_x$ and $t_y$ are the components of the polarization unit vector in
the plane $(\hat{x'},\,\hat{y'})$ which is
perpendicular to \mbox{\boldmath $k$} and has the $\hat{x'}$ axis
along the line of intersection of the $(\hat{x},\,\hat{y})$ and
$(\hat{x'},\,\hat{y'})$ planes.
We keep the previous notation, so the nominal times $\tau_D$ and
$\tau_P$ in Eqs. (A.1)--(A.2),
$$
  \tau_D^{-1}=\frac{\chi^2p_0^3}{2\pi \hbar \rho c_l^2},\quad \tau_P^{-1}=
  \left(\frac{ee_{14}}{\varepsilon}\right)^2\frac{8\pi p_0}
  {5\hbar \rho_0 c_l^2},
$$
have exactly the same magnitudes as they had in
Ref. \onlinecite{di96}. (The notation not explained in the main text
is the deformation potential $\chi$, the piezoelectric constant $e_{14}$,
and the crystal density $\rho_0$.)

If the transverse phonon distribution satisfies the condition that their
polarizations are equiprobable, then averaging of Eq. (A.2) over all
\mbox{\boldmath$t$} directions and subsequent multiplication by 2
to account for
the existence of two transverse polarizations yield
$$
  2\overline{{\tau_{A,{t}}}^{-1}}=\frac{5(c_lp_0)^2}
  {c_t^2 k^6\tau_P}\left(
  q_x^2q_y^2+q^2k_z^2-\frac{9q_x^2q_y^2k_z^2}{k^2}\right)\,.
                                                         \eqno (A.3)
$$

${}\;{}$

\centerline{{\large\bf Appendix II: estimate of the time of
of adiabatic}}
\centerline{{\large\bf equilibrium establishment}}

${}\;{}$

We should check that the time of establishment of
adiabatic equilibrium in 2DEG is shorter than the typical
times (5.11) and (5.18) (since we have used the Bose
distribution (3.3) everywhere). The estimation of this time
$(\Delta t)_{\mbox{\scriptsize ad}}$ may be obtained from
the kinetic relationship $
(\Delta t)^{-1}_{\mbox{\scriptsize ad}}\sim
(N_{\mbox{\scriptsize SW}}/L^2)
\overline{v}_{\mbox{\scriptsize SW}}l$, where
$N_{\mbox{\scriptsize SW}}/L^2$ is the SW density,
$\overline{v}_{\mbox{\scriptsize SW}}=\hbar^{-1}
\overline{{\partial{\epsilon}(q_{\mbox{\scriptsize ex}})/\partial
q_{\mbox{\scriptsize ex}}}}$, which
is the mean SW velocity, and $l\sim
\overline{q}_{\mbox{\scriptsize ex}}l_B^2$, which
is the characteristic cross-section for 2D exciton. Now using
(3.24) and (2.4), and taking into account that
$\overline{q}_{\mbox{\scriptsize ex}}^2l_B^2/2M\sim T$,
we find $(\Delta t)^{-1}_{\mbox{\scriptsize ad}}\sim
MT^2\varrho/\hbar$, where $\varrho=\exp{[(\mu-\delta)/{T}]}$ is
determined in the limiting cases by Eq. (3.25) or by Eq. (5.12)
(according to the magnitude of the temperature $T_0$). One can
see that the double inequality $(\Delta t)_{\mbox{\scriptsize ad}}\ll 
(\Delta t)_T \ll
\tilde{(\Delta t)_T}$ holds. Only in the case of very low temperature
$T_0$ ($T_0<\tilde{T}$) with simultaneously low noneqilibrium
phonon density ($N_s<10^{11}\,$cm${}^{-3}$) do we find
$(\Delta t)_{\mbox{\scriptsize ad}}\gapprox\tilde{(\Delta t)_T}$,
and the presented theory fails. This special region of $T_0$ and
$N_s$ is beyond the scope of our study.

$\;$

$\;$

$\;$

$\;$

$\;$

$\;$

$\;$

$\;$

$\;$

$\;$

$\;$

$\;$

\begin{figure}
\centerline{\psfig{figure=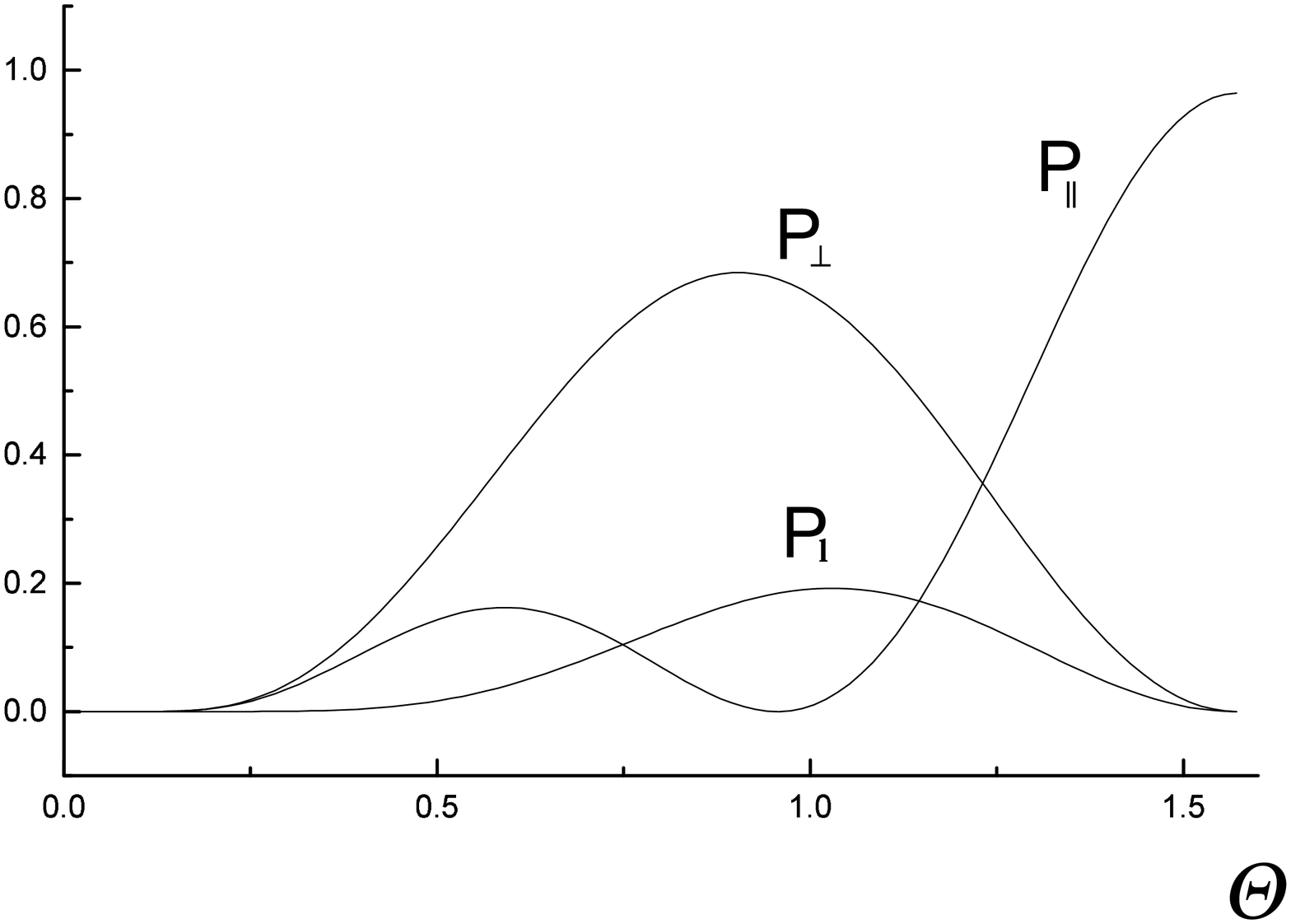,width=120mm,angle=0}}
\caption{Angular dependence of the acoustic wave attenuation coefficients for
various polarizations. These results are obtained for $T=1\,$K,
$B=10\,$T, $1/M=60\,$K; consequently $\delta=3.0\,$K, $\eta_l=0.2$,
$\eta_t=0.07\,$.}

{\noindent
${}\;{}$
\noindent}

\label{Firstfig}
\end{figure}

$\;$

$\;$

$\;$

$\;$

$\;$

\begin{figure}
\centerline{\psfig{figure=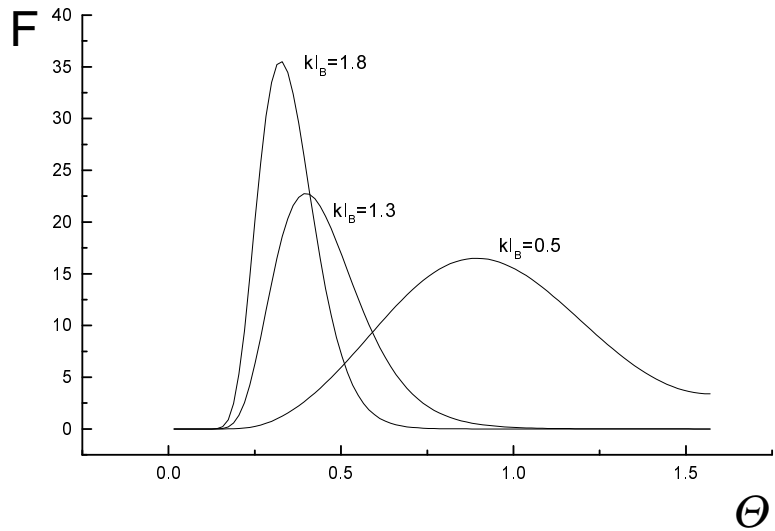,width=120mm,angle=0}}

\caption{The inverse time (3.7) may be represented in the form
$\tau_l^{-1}=L_z^{-1}\exp{[(\mu-\delta)/T]}\cdot$ $F(B,T,k,\theta)$.
In this
figure $F$ is calculated for $B=10\,$T, $T=1\,$K, $\mu-\delta=3\,$K.}

{\noindent
${}\;{}$
\noindent
}
\label{Secondfig}
\end{figure}


\begin{references}
\bibitem{foot1}
{We exclude from the consideration the effects of the heterostructure edges,
  where the gain may be much larger than $\Delta$, and spatial charge
  density fluctuations are possible}.
\bibitem{foot2}
{Note that the screening does not occur in our system due to the absence of
free charges}.
\end{references}
\end{document}